\documentclass{iopconfser}
\usepackage{graphicx}
\usepackage{gensymb}
\usepackage{caption}

\usepackage{subcaption}

\begin{document}

\title{Aerosol Measurement in B, V, and R Filters using Wide-field Photometry}

\author{S Negi$^{1}$, J Ebr$^{1}$, S Karpov$^{1}$, J Eli\'{a}\v{s}ek$^{1,2}$}

\affil{$^1$ FZU -- Institute of Physics of the Czech Academy of Sciences, Na Slovance 1999/2, Prague 182 21, Czech Republic}

\affil{$^2$ Institute of Theoretical Physics, Faculty of Mathematics and Physics, Charles University,\\ V~Hole\v{s}ovi\v{c}k\'ach 2, 180\,00 Prague, Czech Republic}

\email{negi@fzu.cz}

\begin{abstract}
The new method for aerosol measurement using wide-field stellar photometry was originally developed for B-filter data. The dependence of VAOD (vertical aerosol optical depth) on wavelength can be used to understand the physical characteristics of aerosol. The method has been extended for V and R filters using the Gaia stellar photometric catalog as the reference. We calculated the coefficients of the equation describing molecular and aerosol extinction of the atmosphere. For data in a particular filter, they depend on the telescope's location. Light towards shorter wavelengths has higher extinction. The VAOD, calculated in B, V, and R filters does not show monotonous dependence on wavelength, and there is increased dependence on the photometric methods for V and R filters. We studied the possible reasons for this unexpected behavior.
\end{abstract}

\section{Introduction}
FRAMs (F/Photometric Robotic Atmospheric Monitor) are small robotic telescopes used for atmospheric monitoring at CTAO ( Cherenkov Telescope Array Observatory) and Pierre Auger Observatory. These astroparticle-physics experiments indirectly detect ultra-high-energy cosmic rays or very-high-energy gamma photons through the faint (fluorescence or Cherenkov) light created from the extensive air showers as these particles pass through the atmosphere. The transparency of the atmosphere directly affects the intensity of the light created in the air shower.

VAOD (vertical aerosol optical depth) calculation is based on the fundamental principle of light extinction by the Earth's atmosphere. The difference in the measured magnitude to the catalogue magnitude gives the extinction. Previously, we used the Tycho2 catalogue as the reference catalogue, but to calculate VAOD in V and R filters beside the B filter, we used the Gaia DR3 catalogue \cite{2023A&A...674A..33G}. The Angstrom coefficient, which is inversely related to particle size, gives the degree of dependence of the VAOD (vertical aerosol optical depth) on wavelength. The data used in this paper are from the future CTAO (Cherenkov Telescope Array Observatory) \cite{cta} FRAM at Chile. In Chile, two FRAMs are located at a distance of approximately one kilometer to each other; one is referred to as $s0$, and the other as $s1$.

\section{Model data}
\subsection{Modeling extinction coefficient}

The equation (\ref{eqn:model}) compares measured magnitude with the magnitude of the star evaluated by modeling different components of the atmosphere and considering instrumental corrections \cite{VAODpaper}.
\begin{equation}
\label{eqn:model}
\frac{m_\mathrm{inst}}{M} = m_{cat,B}+Z_i+g(A_M,A_A,k_i,B-V)+f(B-V,x,y)
\end{equation}

Here, $m_{inst}$ stands for the measured (instrumental) magnitude, $M$ is the non-linearity correction constant of the photometric process, $m_{cat}$ is the catalogue magnitude using Gaia DR3 catalogue \cite{2023A&A...674A..33G}, $Z_i$ is the photometric calibration constant (known as zero-point), $g$ is the model of extinction for molecular atmosphere and aerosol (considering only tropospheric aerosols) as a function of airmass, star's color index corresponding to the filter in which it is observed, and the value of extinction due to tropospheric aerosols , $f$ accounts for the response of the system to the star's color index and its position $(x,y)$ on the CCD detector.

\begin{figure}[h]
\centering
\includegraphics[width=25pc]{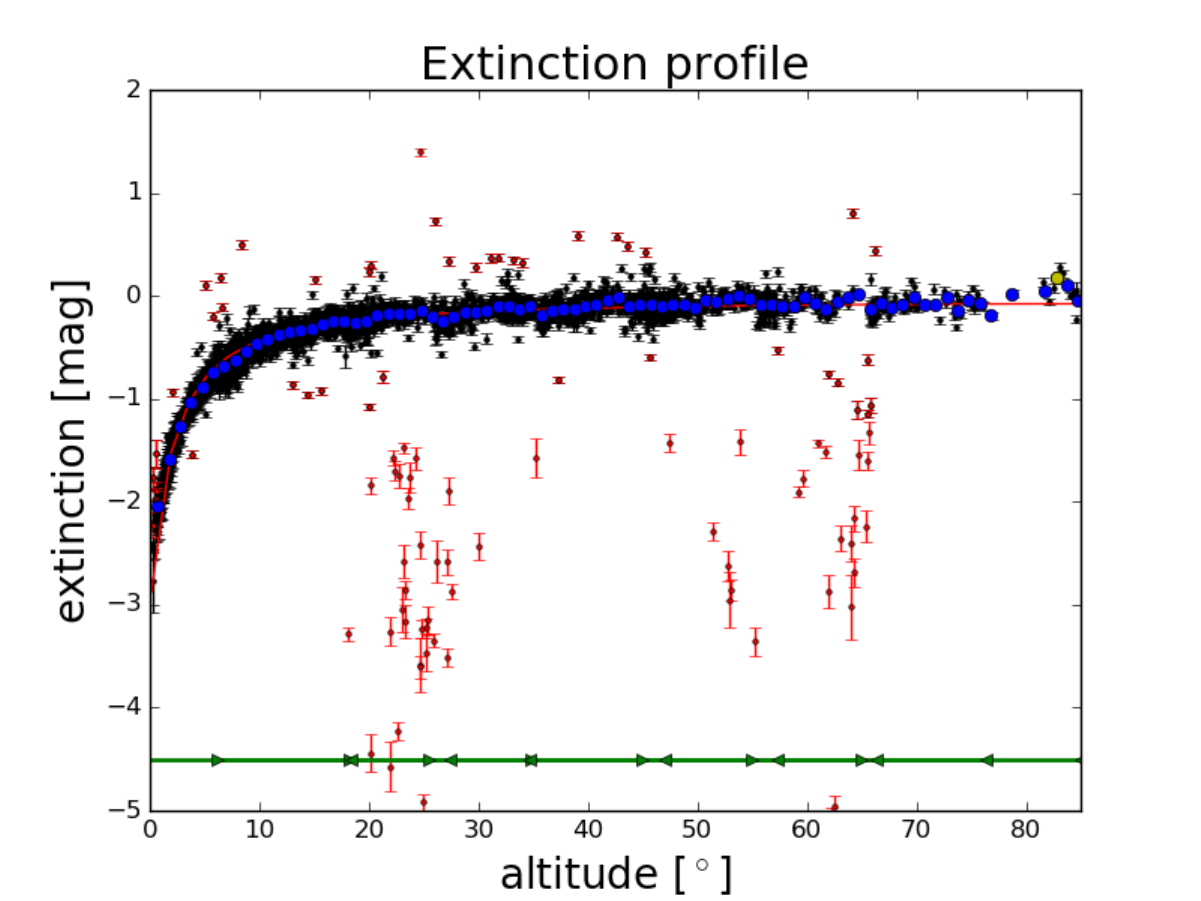}
\caption{\protect\label{mal_nor}The extinction plot of a scan taken in R-band at the CTAO-Chile (s0) site. A scan is a set of 8 images taken at successive altitudes from the horizon to the zenith.}
\end{figure}

The method to calculate VAOD using wide-field stellar photometry was initially developed and optimized for data in the B-filter, using Tycho-2 as the reference catalogue. Now, Gaia DR3 is the reference catalogue for the data processed in B, V, and R filters. To avoid light contamination and thus perform reliable photometry of the stars, we reject stars with any source in its 20-pixel radius to avoid light contamination by other sources for correct magnitude estimation. Processing of the data in the R-band showed that the number of stars detected is higher for the approximately same field in the B-filter, causing the rejection of a significant number of stars, as seen in the Fig.~\ref{mal_nor} at the altitudes above 30\degree of the extinction plot.

\begin{figure}
\centering
\begin{subfigure}{.4\textwidth}

  \includegraphics[width=1.\linewidth]{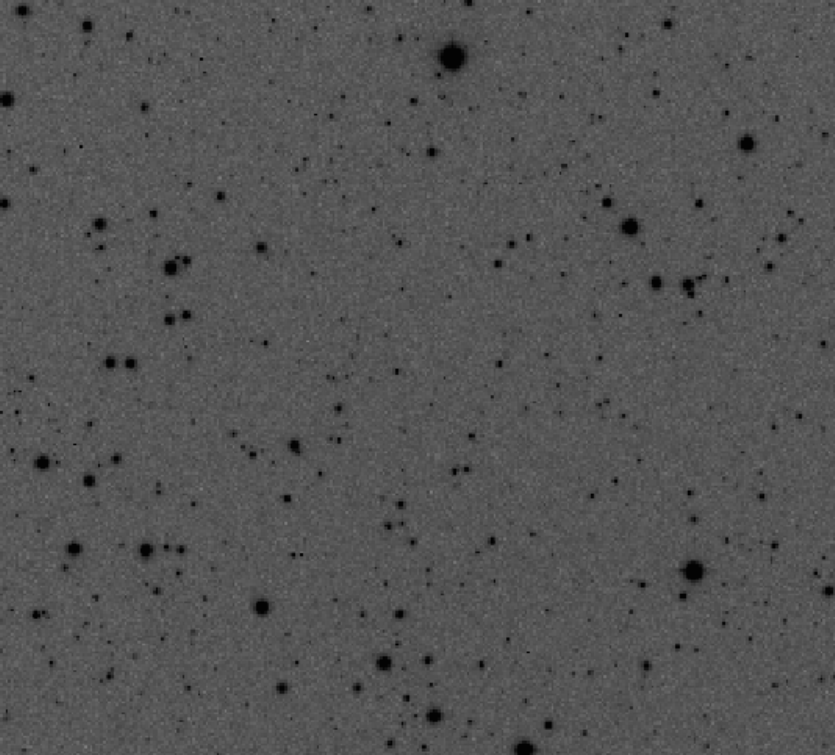}
  \caption{AT altitude 17\degree}
  \label{fig:2sub1}
\end{subfigure}\hspace{5mm}%
\begin{subfigure}{.4\textwidth}
  
  \includegraphics[width=1.\linewidth]{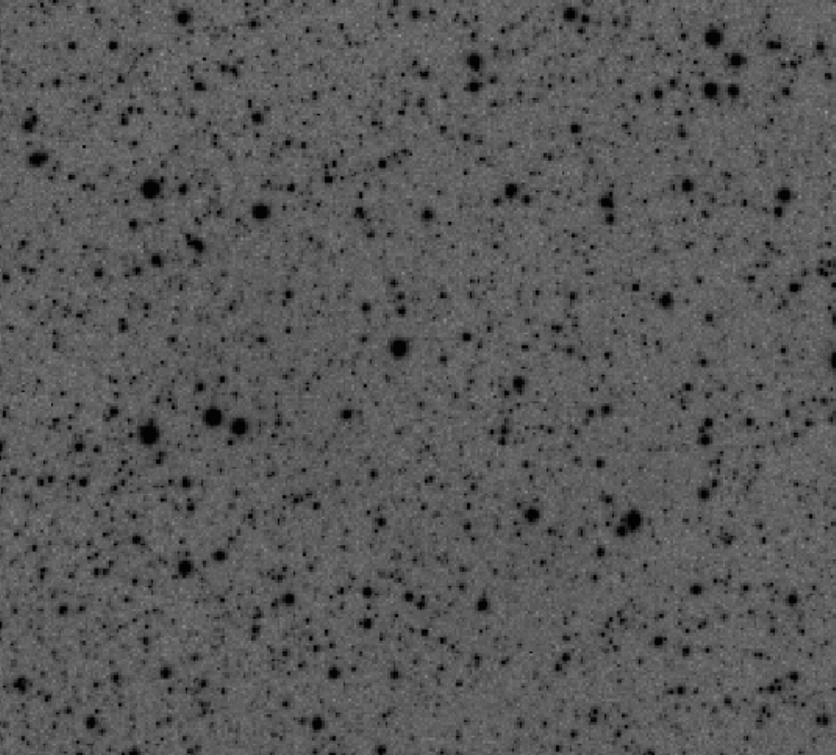}
  \caption{At altitude 77\degree}
  \label{fig:2sub2}
 
\end{subfigure}
\caption{Both images belongs to the scan in the Fig.~\ref{mal_nor} in R-band. Fig.~\ref{fig:2sub1} is at altitude approx. 17\degree and Fig.~\ref{fig:2sub2} is at the altitude approx. 77\degree}
\label{fig:fig2}
\end{figure}

\begin{figure}
\centering
\begin{subfigure}{.4\textwidth}

  \includegraphics[width=1.\linewidth]{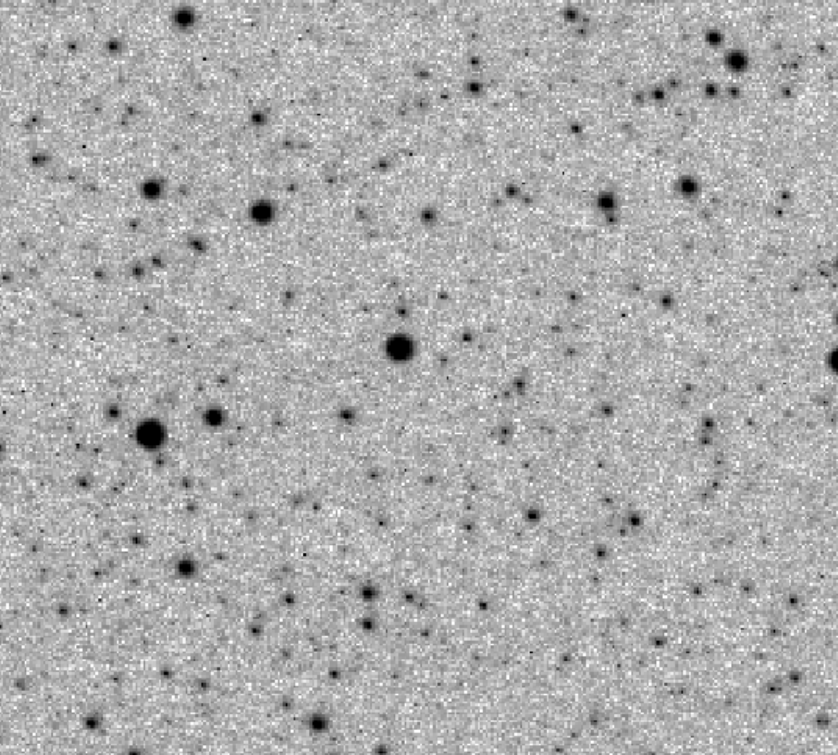}
  \caption{In B-band}
  \label{fig:3sub1}
\end{subfigure}\hspace{5mm}%
\begin{subfigure}{.4\textwidth}
  
  \includegraphics[width=1.\linewidth]{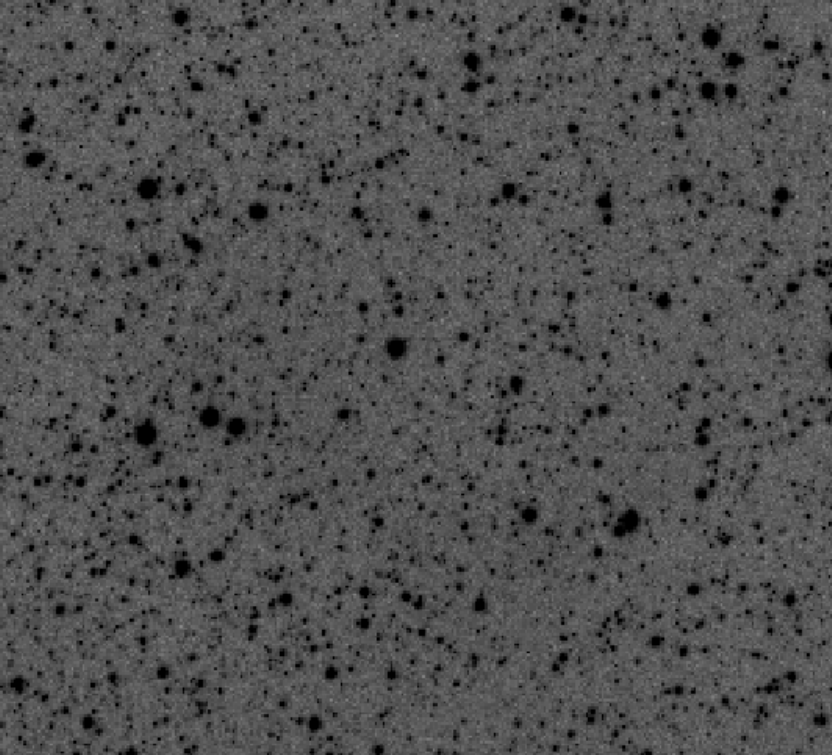}
  \caption{In R-band}
  \label{fig:3sub2}
 
\end{subfigure}
\caption{Both the images were taken at the same altitude and azimuth with a difference of 2 minutes and 7 seconds. The Fig.~\ref{fig:3sub2} is same as Fig.~\ref{fig:2sub2}}
\label{fig:fig3}
\end{figure}

We analyzed the images for such scans and found that crowded fields caused higher rejection of stars. As seen in the Fig.~\ref{fig:2sub1}, the field is less crowded than the field in the Fig.~\ref{fig:2sub2}, so more stars are detected. Also, when we compared the image in the Fig.~\ref{fig:3sub2} in R-band with the image in Fig.~\ref{fig:3sub2} (which is same as the Fig.~\ref{fig:2sub2}) taken at the same altitude and azimuth in the B-band at a time difference of 2 minutes and 7 seconds, we found that the same approximate field was more crowded in the R-band than the B-band. Thus, we reduced the minimum distance between the sources to 10 pixels for rejection.

\subsection{Extinction in magnitude}

For a beam of light at a particular wavelength, the extinction in magnitude as a function of airmass $A$ is expressed as

\begin{equation}
\label{eqn:extinction}
\Delta m = -2.5 log_{10}(exp(\tau(\lambda)A)) = \frac{2.5\tau(\lambda)A}{ln_{10}} \approx 1.086\tau(\lambda)A = k(\lambda)A
\end{equation}

\begin{equation}
\label{eqn:tau}
k = 1.086\tau
\end{equation}

where, k is the extinction coefficient and $\tau$ is optical depth.

The equation (\ref{eqn:model}) is fitted for the value of $Z_i$ (zero-point) and $k_i$ extinction coefficient of tropospheric aerosols in magnitude, and then the (\ref{eqn:tau}) was used to determined VAOD for each scan.  
\section{Photometric method}

We calculate the measured magnitude of the star to compare it with the modeled magnitude as in the equation (\ref{eqn:model}). We have implemented IRAF (Image Reduction and Analysis Facility) \cite{iraf} for aperture photometry. In aperture photometry,  all the pixels within a defined aperture (in pixels) are used to calculate the star's magnitude. To find the optimum aperture value, we calculated the RMS value of the difference between the measured magnitude and the modeled magnitude of all the stars in the same set of (forty) scans nine times with an aperture ranging from one to nine pixels. We repeated this for the two sites (CTAO-s0 $\&$ s1) in the B, V, and R filters. The aperture at which this value is minimum is the optimum aperture; it was two pixels for both the sites in all the filters. According to the paper \cite{VAODpaper}, the optimum aperture was four pixels. We inferred that its value depends on the data set used for the evaluation.

\begin{figure}[h]
\includegraphics[width=37pc]{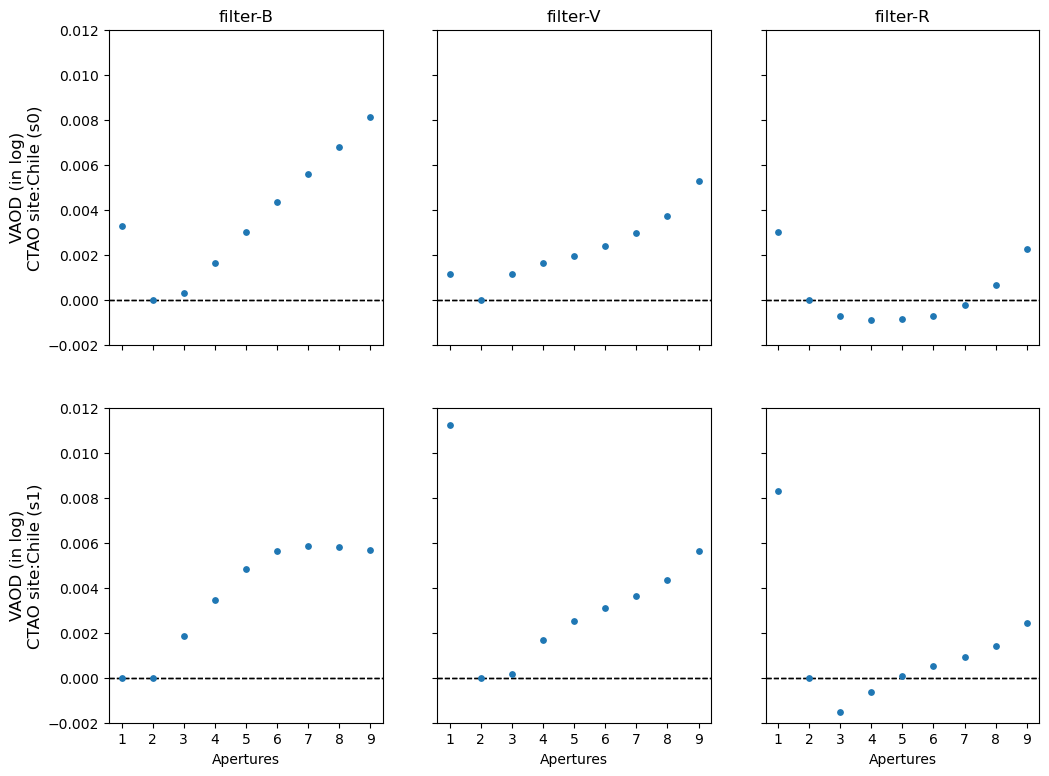}
\caption{Each plot represents the difference of VAOD calculated at aperture values (in pixels) from 1 to 9 for aperture photometry with the VAOD at aperture equal to 2 pixels. The first and second rows show CTAO at Chile (s0, s1), respectively. Furthermore, the columns from left to right represent filters B, V, and R.}
\label{fig:vaod_diff}
\end{figure}
The Fig.~\ref{fig:vaod_diff} represents the dependence of the calculated VAOD value on the aperture size, which is higher than 0.003, as estimated in the previous publication \cite{VAODpaper}, while processing data only in B-filter.

\section{Angstrom coefficient}
The optical depth of aerosol depends on the wavelength of light according to the power law as
\begin{equation}
\label{eqn:opticaldepth}
\tau(\lambda) = \tau_0 ({\lambda}/{\lambda_0})^\alpha
\end{equation}

VAOD value for B, V and R filter is used in three different combinations to determine $\alpha$
\begin{equation}
\label{eqn:angstrom}
\alpha = -\frac{\log_{10}(\tau_{\lambda1}/\tau_{\lambda2})}{\log_{10}(\lambda1/\lambda2)}
\end{equation}
We calculated VAOD in B, V, and R filters and used 440 nm, 540 nm, and 620 nm,  which refer to the wavelengths at maximum transmission for the respective filter. Moreover, it is used in the equation (\ref{eqn:angstrom}) in a combination of B-V, V-R, and B-R to evaluate the Angstrom coefficient's value.

\section{Results}
\begin{figure}[h]
\includegraphics[width=37pc]{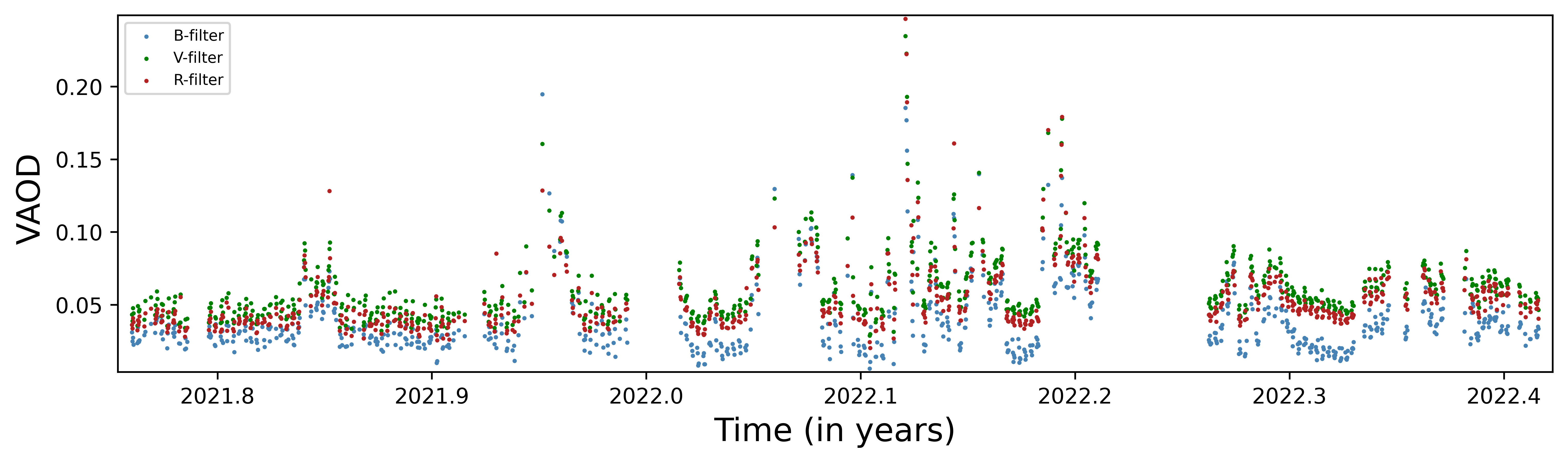}
\caption{\protect\label{fig:VAOD}The plot represents the time-lapse of the VAOD for site CTAO (s1). This data is used to calculate the Angstrom coefficient.}
\end{figure}

\begin{figure}[h]
\includegraphics[width=37pc]{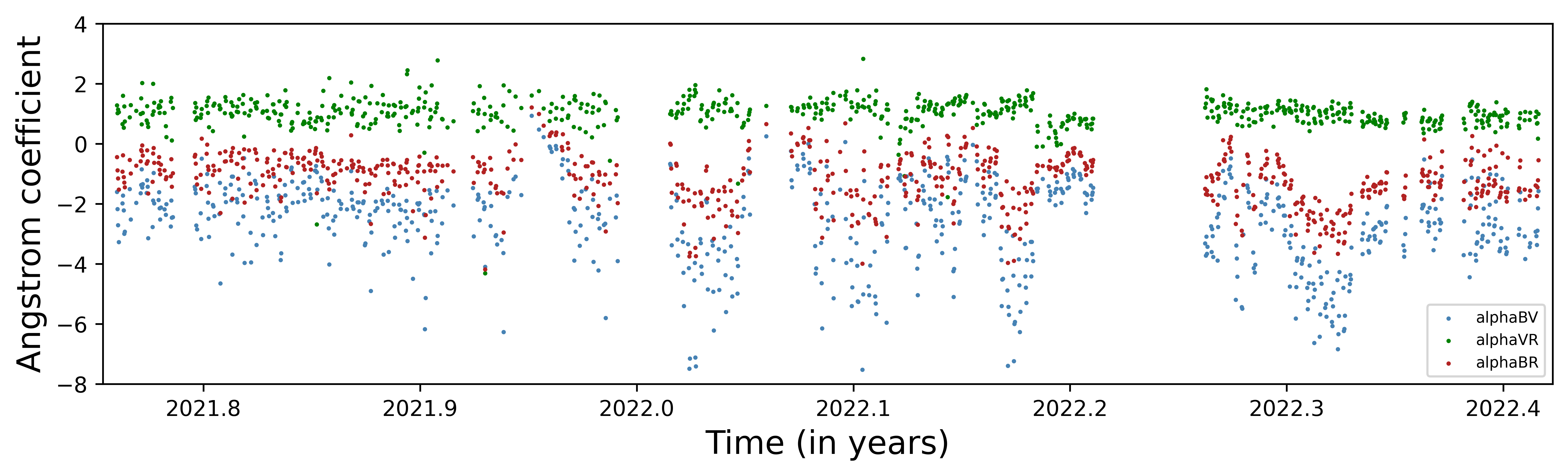}
\caption{\protect\label{fig:BVR}Plot of the time-lapse of the Angstrom coefficient for site CTAO (s1). In the Fig.~\ref{fig:BVR}, the points in blue refer to VAOD and wavelength in the B and V filters as a combination to calculate the Angstrom coefficient. Similarly, points in green refer to a combination of V and R filters, and points in red to B and R filters.}
\end{figure}

We used the three combinations in the pair of two of the VAOD values using the existing method in the three filters to evaluate the Angstrom coefficient in the equation (\ref{eqn:angstrom}). In principle, each combination should evaluate the approximate exact value of the Angstrom coefficient. However, this is not the case for some unknown reason, as seen in the Fig.~\ref{fig:BVR}, and it is the same for the data processed for other sites. Furthermore, the VAOD value for the same scan in the three filters does not always follow the B$>$V$>$R pattern, which causes a negative Angstrom coefficient value. The data presented in Fig.~\ref{fig:VAOD} and Fig.~\ref{fig:BVR} is from the CTAO site (s1) because no changes were made in the photometric system during the time period presented in the Fig.~\ref{fig:BVR}. It contains approximately eight months of data, including the period of the eruption of the Hunga Tonga volcano in December 2021. Since 2022.0 in the above figure, there is more deviation in the Angstrom value for the three combinations, the reason for which is unknown. The above-summarised information is limited to the set of data processed and analyzed.

\section{Conclusion}
 
The Angstrom coefficient calculated by the three combinations of VAOD value in the B, V, and R filters (B-V, V-R, R-B) is not the same, which holds for all the data sets processed for all three FRAMs at future CTAO sites. Furthermore, this value is mostly negative for the period presented in the Fig.~\ref{fig:BVR} for the CTAO site (s1). The reason for this behavior is unknown. 

Since the VAOD value at the CTAO site in Chile is small, we suspect that the method is not sensitive enough to measure the Angstrom coefficient for low VAOD values. We are also testing the data for technical issues. Thus, we are thoroughly inspecting the processed data for each CTAO site to see if any change in the equipment is affecting the final VAOD value.
\section*{Acknowledgement}

This work was supported by the Ministry of Education, Youth and Sports, MEYS  LM2018105, LM2023047 and EU/MEYS CZ.02.01.01/00/22\_008/0004632. This work was conducted in the context of the CTA Consortium.

\bibliography{ebr-atmohead22}
\bibliographystyle{iopart-num}
\end{document}